\documentclass[twocolumn,english,aps,pra,superscriptaddress,amsmath,amssymb,floatfix,nofootinbib,longbibliography
]{revtex4-2}

\usepackage{amsthm}
\usepackage{amsfonts}
\usepackage{siunitx}
\usepackage{amsmath}
\usepackage{amssymb}
\usepackage{graphicx}
\usepackage{verbatim}
\usepackage[colorlinks]{hyperref}
\usepackage{tikz}
\usepackage{pgfplots}
\usepackage{adjustbox}
\usepackage{braket}
\usepackage{xcolor}
\usepackage{physics}
\usepackage{amssymb} 
\usepackage{graphicx}
\usepackage{dcolumn}
\usepackage{bm}
\usepackage{mathtools}
\usepackage{hyperref}
\usepackage{mathrsfs}
\usepackage{dashrule}
\usepackage{caption}
\usepackage{subcaption}
\usepackage{quantikz}
\usepackage[version=4,arrows=pgf-filled,
textfontname=sffamily,
mathfontname=mathsf]{mhchem}

\usepackage[font=small,labelfont=bf,
   justification=justified,
   format=plain]{caption}

\definecolor{linkcolor}{RGB}{0,83,166}
\hypersetup{
  colorlinks = true,
  allcolors = {linkcolor}
}

\begin{document}

\title{Numerical Experiments with Parameter Setting of Trotterized Quantum Phase Estimation for Quantum Hamiltonian Ground State Computation}

\author{Elijah Pelofske}
\email[]{epelofske@lanl.gov}
\affiliation{Information Systems \& Modeling, Los Alamos National Laboratory, Los Alamos, NM, USA}
\affiliation{Quantum \& Condensed Matter Physics, Los Alamos National Laboratory, Los Alamos, NM, USA}
\affiliation{Center for Quantum Computing, Los Alamos National Laboratory, Los Alamos, NM, USA}

\author{Stephan Eidenbenz}
\email[]{eidenben@lanl.gov}
\affiliation{Information Sciences, Los Alamos National Laboratory, Los Alamos, NM, USA}
\affiliation{Center for Quantum Computing, Los Alamos National Laboratory, Los Alamos, NM, USA}

\begin{abstract}
We numerically investigate quantum circuit elementary-gate level instantiations of the standard Quantum Phase Estimation (QPE) algorithm for the task of computing the ground-state energy of a quantum magnet; the disordered fully-connected quantum Heisenberg spin glass model. We consider (classical simulations of) QPE circuit computations on relatively small quantum Hamiltonians ($3$ qubits) with up to $10$ phase bits of precision, using up to Trotter order $10$. We systematically study the inputs of QPE, specifically time evolution, Trotter order, Trotter steps, and initial state, and illustrate how these inputs practically determine how QPE operates. From this we outline a coherent set of quantum algorithm input and tuning guidelines. One of the notable properties we characterize is that QPE sampling of the optimal digitized phase converges to a fixed rate. This results in strong diminishing returns of optimal phase sampling rates which can occur when the Trotter error is surprisingly high.

\end{abstract}

\maketitle

\section{Introduction}
\label{section:Introduction}

In this study we address the task of ground-state energy estimation of non-commuting many-body quantum Hamiltonians, using numerical simulations of quantum circuit implementations and algorithmic engineering of the Quantum Phase Estimation (QPE) algorithm. Quantum Phase Estimation~\cite{kitaev1995quantummeasurementsabelianstabilizer, nielsen2010quantum} is the core technique that famous quantum algorithms such as Shor's algorithm~\cite{365700, Shor_1997}, quantum counting~\cite{Brassard_1998}, and HHL~\cite{PhysRevLett.103.150502} make use of to speed up computations compared to classical algorithms. QPE can additionally be used for the task of minimum eigenvalue finding, which in turn can be used to find the ground-state of a quantum Hamiltonian~\cite{Abrams_1999}. This type of quantum computation, digital minimum eigenvalue finding and quantum simulation, is one of the most promising applications of quantum computers. However, QPE specifically requires immense circuit depths and relatively high qubit counts, making it intractable to apply to current noisy quantum computers. Nonetheless, QPE resource estimate scaling is an important consideration for eventually running full large scale quantum computations on error-corrected quantum computers. Moreover, circuit level instantiations of QPE, and in particular full classical statevector validation of the QPE algorithm performance (this is limited to $\approx 35$ qubits, due to the exponential scaling of exact quantum mechanical simulation using classical numerical methods), remains relatively unexplored. Prior studies have considered some numerical simulations on Fermionic-based molecular Hamiltonians with some quantum circuit model implementations~\cite{johnstun2021optimizingphaseestimationalgorithm, nelson2024assessmentquantumphaseestimation, Ding_2023, patel2024optimalcoherentquantumphase, apel2025reducingquantumresourcesobservable}, but in general there is a lack of rigorous algorithmic engineering analysis and algorithm implementation testing of QPE on specific Hamiltonian instances in the literature. There are many aspects of the algorithm including the initial state, and most importantly how the controlled unitary evolution is implemented, that dramatically impact how well QPE works even for very small quantum Hamiltonian instances. Ultimately, QPE uses a controlled simulation of the unitary $U = e^{i H t}$ as the core algorithmic primitive that is very computationally intensive that allows the computation of the ground-state energy of the quantum Hamiltonian $\mathcal{H}$. In this work, we consider Trotter-Suzuki decomposition~\cite{trotter1959product, suzuki1976generalized, suzuki1985decomposition, Hatano_2005, lopezcerezo2025rigorousintroductionhamiltoniansimulation, Berry_2006} in order to implement this unitary evolution within QPE~\cite{Kivlichan_2020}, but other methods exist such as qubitization~\cite{Low_2019, Berry_2019}.

We illustrate the intrinsic details of QPE by carrying out small-scale (small Hamiltonian) numerical experiments to probe QPE algorithm implementations, all with the goal of computing the ground-state energy of a quantum Hamiltonian. Specifically, we implement and test the standard, canonical, QPE algorithm, sometimes called ``textbook QPE''. The quantum magnet model that we consider is a general Heisenberg spin glass, with no local fields, defined as
\begin{equation}
\label{eq:heisenberg_glass}
\mathcal{H}
\,=\,
\sum_{i<j} 
\Big(
  J^{x}_{ij}\,\sigma_i^{x}\sigma_j^{x}
 +J^{y}_{ij}\,\sigma_i^{y}\sigma_j^{y}
 +J^{z}_{ij}\,\sigma_i^{z}\sigma_j^{z}
\Big),
\end{equation}
where each $J^{\alpha}_{ij}$ coefficient is chosen independently and uniformly at random to be $\pm 1$, and \(\sigma_i^{\alpha}\) (\(\alpha = x,y,z\)) are Pauli operators at each site. This model is a type of disordered anisotropic Heisenberg quantum Hamiltonian~\cite{Itoi_2024, bray1980replica, Georges_2000}, with long range interactions similar to the Sherrington-Kirkpatrick model~\cite{PhysRevLett.35.1792, de_Almeida_1978}. This Hamiltonian is a good candidate for QPE numerical simulations for quantum ground-state finding because it has non-trivial magnetic frustration, degeneracy, and is a canonical example of a general non-commuting many-body quantum Hamiltonian. Ground-state computation of Heisenberg models enables the understanding of low temperature properties of materials, especially frustrated magnetic materials~\cite{dannegger2025quantumfluctuationsdeterminespinflop, Shinaoka_2011, PhysRevB.103.205122, PhysRevB.107.224411, Torelli_2020, v2fm-zk8l, PhysRevB.105.085120}. Moreover, low-temperature computation of quantum Heisenberg model Hamiltonians is a widely used technique in computational physics, typically using exact or approximate diagonalization, tensor network methods such as Density Matrix Renormalization Group (DMRG), and Quantum Monte Carlo~\cite{PhysRevLett.133.016501, Tsai_2020, PhysRevB.23.6126, PhysRevB.98.134427, PhysRevX.11.041021, uematsu2021frustration, PhysRevB.92.134407, han2024entropydynamicsbinarybond}. Therefore, it is of interest to be able to reliably compute the ground-state of quantum Heisenberg models. Quantum Monte Carlo suffers (QMC) from the famous sign problem when approximating the ground-state of certain types of quantum Hamiltonians such as Heisenberg spin glass models~\cite{Pan_2024, Iglovikov_2015}. DMRG is the other alternative classical algorithm for these types of computations, but it fails to numerically converge on large highly connected spin systems, or when the entanglement is very high. Quantum Phase Estimation could therefore be a viable future more efficient alternative to both DMRG and QMC for these types of computations.

We do not address the question of determining the magnetic phase of this model; e.g., whether this is a quantum spin glass phase. Instead, we study this model strictly because it is a highly frustrated, highly degenerate, quantum magnet model: Computing the ground-state of Heisenberg spin glasses would be a substantial capability, and in this study we aim to investigate the algorithmic capabilities -- via numerical scaling tests -- of the Quantum Phase Estimation (QPE) algorithm, where the unitary is implemented approximately by Trotterization, a type of product formula. A significant amount of research has gone into QPE on Fermionic systems~\cite{tranter2025highprecisionquantumphaseestimation, kang2022optimizedquantumphaseestimation, Whitfield_2011, Aspuru_Guzik_2005, Ino_2024}. This study considers distinct spin-based condensed-matter physics Hamiltonians.

\begin{figure*}[ht!]
\centering
\begin{quantikz}[row sep=0.25cm, column sep=0.4cm]
\lstick[wires=3]{$\text{phase bitstring readout register}$}
    & \ket{0} & \gate{H}
    & \qw
    & \qw
    & \ctrl{3}
    & \gate[wires=3]{\mathrm{QFT}^{-1}}
    & \meter{} \\[-0.2em]
& \ket{0} & \gate{H}
    & \qw
    & \ctrl{2}
    & \qw
    &
    & \meter{} \\ [-0.2em]
& \ket{0} & \gate{H}
    & \ctrl{1}
    & \qw
    & \qw
    &
    & \meter{} \\ \lstick[wires=3]{$\text{unitary evolution register}$}
& \ket{0} & \gate[wires=3]{\mathcal A}
    & \gate[wires=3]{U^{2^{0}}}
    & \gate[wires=3]{U^{2^{1}}}
    & \gate[wires=3]{U^{2^{2}}}
    & \qw
    & \qw \\& \ket{0} & \qw
    & \qw
    & \qw
    & \qw
    & \qw
    & \qw \\&\ket{0} & \qw
    & \qw
    & \qw
    & \qw
    & \qw
    & \qw
\end{quantikz}
\caption{\textbf{Quantum circuit diagram schematic of the Quantum Phase Estimation algorithm to compute the minimum eigenvalue of a quantum Hamiltonian, using $3$ qubits for the phase register and $3$ qubits for the unitary evolution register as a small scale example. } The unitary evolution register is initialized in the state ${\mathcal A}$, followed by evolution by controlled powers of $U$, which in this case are implemented by Trotter-Suzuki (or first order Lie-Trotter) decomposition. The algorithm terminates by performing an inverse quantum Fourier transform (QFT) on the phase qubits, at which point the qubits are measured in the computational $Z$ basis. A single measured bit vector from the algorithm is an estimate of the phase, which can be translated to an estimate of the ground-state energy of a quantum Hamiltonian. Here, $H$ denotes the Hadamard gate.  }
\label{figure:QPE_circuit}
\end{figure*}
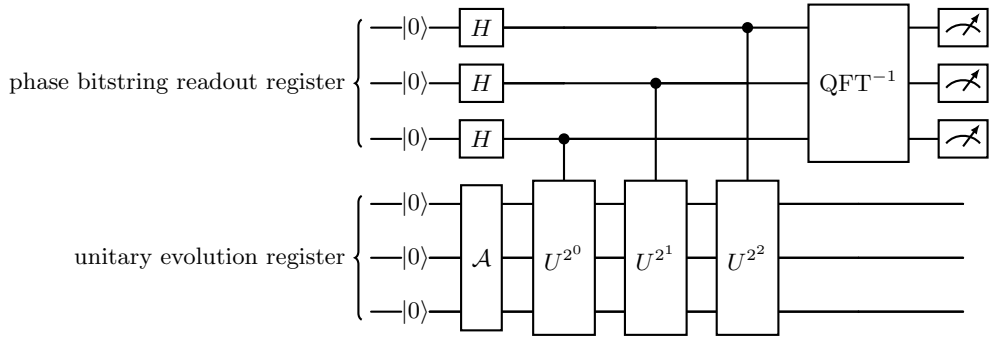

The aim of this study is to provide parameter setting reasoning, at a scale that is exactly numerically tractable with classical computation, of QPE for the purpose of informing future use and implementation on digital quantum computers. QPE is widely viewed as one of the most consequential quantum algorithms, but it can only be executed by full scale error corrected quantum processors, which are not yet physically realized. To this end, QPE will require algorithms engineering understanding to be developed, and here we consider a small scale test case of this by means of numerical execution of explicit descriptions of QPE digital circuits.

\section{Methods}
\label{section:methods}

Algorithmically, QPE is defined by several key components; we describe specifically the Trotterization implementation of QPE. First, we want the ground-state energy (which we will typically notate as $E_0$) to some number of bits of precision, we will call this $m_{prec}$ (sometimes this is also called the number of ancilla qubits). The measured binary phase that we get is $\phi = 0.\phi_1 \phi_2 \phi_3 \ldots \phi_{m_{prec}}^{(2)}$ where each $\phi_i$ is a classical bit. More bits of precision means that we can compute, assuming the rest of the algorithm is properly tuned, a better digitized estimate of the optimal ground-state energy of a quantum Hamiltonian. For a given $m_{prec}$ and a total evolution time, there is a single bitstring which corresponds to the optimal digitized phase \footnote{We do not consider the somewhat pathological case where the true eigenvalue is exactly between two bitstrings. }. The optimal phase $\phi$ can then be translated to the energy of the quantum Hamiltonian by $\Tilde{E} = - \frac{2 \pi}{t} 0.\phi_1 \phi_2 \phi_3 \ldots \phi_{m_{prec}}^{(2)}$, where ``nearest bin decoding'' is used. An example $3$-qubit unitary and $3$-bit phase readout register quantum circuit diagram is shown in Fig.~\ref{figure:QPE_circuit}, where the unitary simulation blocks are controlled on each of the phase qubits. The remaining QPE algorithmic components are parameters that can be tuned, given a fixed $m_{prec}$, outlined next.

\emph{Evolution time $t$} is an incredibly important parameter for this computation; $t$ needs to be sufficiently large that the lower energy eigenvalues can be clearly distinguished from each other, such as between the ground state and the first excited state. However, $t$ can cause aliasing, or phase wrap, when it is too large which introduces ambiguity into how eigenvalue estimates are computed with QPE. Specifically, phase wrapping without identifying how many times phase wrapping has occurred results in the decoded energy estimates converging closer to zero (and therefore being incorrect). Identifying when phase wrapping has occurred is hard in general. There are no established techniques, as far as we are aware, to identify phase wrapping within QPE quantum circuit sampling. For arbitrary quantum Hamiltonians the time evolution can be set using any upper bound on the spectral range of the Hamiltonian: $t_{max} = \frac{\pi}{|| H ||_{\infty}}$. For this study, we start with a $t$ given by a generic bound that is efficient to compute and is used in the Qiskit implementation~\cite{javadiabhari2024quantumcomputingqiskit, gadi_aleksandrowicz_2019_2562111}; 

\begin{equation}
    \label{equation:heuristic_evolution_time}
    t_0 = \frac{\pi}{3 |E| |J|}. 
\end{equation}

Here, $|J|=1$ for the Heisenberg model, and $|E|$ is the number of edges in the interaction graph (induced by the Pauli strings of the Hamiltonian), which for our $n$-node Hamiltonian is $|E| = \frac{n\cdot(n-1)}{2}$ because the graph is a clique. While this is a reasonable, and efficiently computable bound (aliasing does not occur when using this bound), changing $t$ is one of the tunable parameters that significantly impacts the performance of a QPE implementation -- as we will show. The value of $t$ is linked to the number of bits of precision $m_{prec}$ as follows: when $m_{prec}$ increases, energy level differentiation becomes harder, in particular measurements of two energy states (such as the ground-state and the first excited state) become indistinguishable if $t$ is not sufficiently large. Moreover, each $t$ corresponds to a single optimal phase bitstring. In other words, the tuning of $t$ is linked to how large $m_{prec}$ is, and when $m_{prec}$ is large $t$ must also be large (without causing phase wrapping).

The \emph{number of Trotter steps}, which we will denote as $r$, is the third key parameter. A higher number of Trotter steps means that the unitary evolution is better approximated and thus results in smaller Trotter error. For this study, we begin the number of Trotter steps at $1$ and increase up to either convergence or until it becomes infeasible to execute the circuits. 

\emph{Trotterization order}, which we will denote as $k$, is the fourth parameter. The unitary time evolution approximation we use is the Trotter-Suzuki decomposition of orders 1, 2, 4, 6, 8, and 10. Higher order Trotterization results in lower Trotter error, but also comes with an increased computational cost due to an increase in total circuit instructions. The primary way that Trotter order and Trotter steps interact with the evolution time is that longer evolution times require lower error Trotterization to correctly approximate the time dynamics.

\emph{The initial state of QPE} is the fifth and final algorithmic consideration; the initial state for the unitary evolution must have a non-zero overlap with an eigenstate of the unitary operator we are simulating in order for the algorithm to work. In this study, we evaluate several different initial state choices of QPE, including both product states and short-depth entangled states. We consider the following states as potential initial states; 1) the all zero state, 2) an $n$-qubit graph clique state, 3) an $n$-qubit GHZ state, 4) a $n$-qubit random SU(4) Quantum Volume circuit~\cite{Cross_2019, aaronson2016complexitytheoreticfoundationsquantumsupremacy} with random parameters, 5) randomly parameterized (with continuous parameters) U3 single qubit gates on each qubit, and 6) staggered X gates. The initial state overlap quantity, taking into account eigenvalue degeneracy, is defined as 

\begin{equation}
    \label{equation:degenerate_overlap}
    \chi = \sum_{\psi_2 \in v_0} \left| \langle \psi_1 \mid \psi_2 \rangle \right|^2, 
\end{equation}

where $\psi_1$ is the initial state, and $v_0$ is the set of eigenvectors that have the same minimum eigenvalue $E_0$ each $\psi_2$ is an eigenstate corresponding to a ground state $\ket{\psi_2}$. Degenerate eigenvalues are determined using a numerical precision threshold of $1 \times 10^{-12}$, and the set $v_0$ is comprised of orthonormal eigenvectors.

\paragraph*{QPE implementation and numerical details} In summary, the inputs we supply to (Trotterized) QPE for ground-state computation are the target Hamiltonian $\mathcal{H}$, the initial state $\mathcal{A}$, Trotter order $k$, Trotter steps $r$, and evolution time $t$. The numerical experiments that we perform are to probe the performance of QPE when these four key QPE parameters are tuned, given a fixed Hamiltonian and fixed bit phase readout register size. One final consideration is that this computation is inherently probabilistic, and in particular we must sample a distribution of measured bitstrings, which are estimates of the optimal phase. For this study, we use a fixed sample count of $10{,}000$ measurements for each QPE parameter combination. One of the central questions that we examine is how to run QPE in the context of it being a sampling algorithm; decoding an ensemble of measured bitstring energies and then taking the minimum energy does not guarantee that is a true ground-state of the quantum Hamiltonian of interest, due to a variety of factors within QPE including primarily digitization error. To this end, we will primarily measure the optimal ground state energy bitstring probability.

We analyze the QPE implementation accuracy using the ground-truth of full exact diagonalization to determine the true ground-state energy $E_0$. This is computed specifically using the Python 3 library Numpy~\cite{harris2020array}, which in turn uses the LAPACK linear algebra library~\cite{129995}. The fundamental machine precision limit for this exact diagonalization routine is $\approx 1 \times 10^{-16}$. The QPE implementation we use is Python 3 based and uses modified versions of various existing software library implementations of QPE, mostly based on Qiskit~\cite{javadiabhari2024quantumcomputingqiskit}, with more details given in Appendix~\ref{section:appendix_QPE_circuit_implementation}. The source code uses time exponentiation for each phase qubit and \texttt{complex128} datatypes. Qiskit version \texttt{2.2.2} was used for all quantum circuit description and quantum circuit sampling computations~\cite{javadiabhari2024quantumcomputingqiskit}. The circuit simulations were run with reasonably time-intensive HPC (High Performance Computing) resources. One aspect of the QPE implementation which we do not analyze or optimize is the ordering of operator exponentials in the Trotter product formula. For a given Hamiltonian, $r$, and $k$, there exist multiple unique Trotter orderings. Different operator ordering results in different effective Hamiltonians, and therefore can have different effective errors (how well it approximates the target unitary), as well as different circuit level descriptions~\cite{Childs_2021, Childs_2019, Tranter_2019, hastings2014improvingquantumalgorithmsquantum}. Additionally, we apply the same (controlled) Trotter decomposition defined by $r, k$ on each phase qubit. In this case, we use the default Trotterization implemented in Qiskit, which does allow reordering of terms to minimize circuit depth. The QPE circuits are decomposed into elementary single two-qubit gate instructions to minimize circuit depth using a heuristic graph coloring of a commutator graph, however, the circuits are not optimized, compiled, or routed, so as to preserve the original circuit structure without the potential for unintentional approximation errors. Importantly, the QPE implementation does not make use of any information about the problem Hamiltonian, other than naive bounds to set the total evolution time; the goal is to investigate how to apply QPE in a ``black-box'' algorithm manner. In particular, setting of the total simulation time is not informed by minimum eigenvalue information from exact diagonalization. Exact diagonalization is used to validate the sampling of the QPE circuit, but not used in the construction of the circuit or the choice of algorithm parameters.

\paragraph*{Ideal initial state digitized phase sampling rate} Another important quantity for analyzing QPE is the effective optimal phase QPE upper bound, which is given by ref.~\cite{kaye2006introduction} in Lemma 7.1.2, reproduced here with the notation used in our manuscript:

\begin{equation}
    \label{equation:px_upper_bound}
    p(x) = \frac{1}{2^{2m}} \frac{ \sin^2(\pi(2^m \phi - x)) } {\sin^2 (\pi (\phi - x/2^m))}, 
\end{equation}

where $\phi$ is the non-digitized phase of $E_0$ that is computed using the true smallest eigenvalue of the Hamiltonian (and some $t$), $m$ is the number of phase qubits, and $x$ is the integer decimal representation of the digitized phase $\phi$ (ranging from 0 to $2^m-1$). This bound takes as input (at least, assuming we care about $p(x)$ for the optimal phase) only the size of the Hamiltonian, $n$, the number of phase bits of precision, the phase, and implicitly the total time evolution because the optimal $\phi$ depends on $t$. Eq.~\ref{equation:px_upper_bound} assumes the perfect possible initial state, which not only can we not prepare, being able to prepare it would solve the very problem we wish to use QPE to solve. Therefore, our initial state overlap will always be worse than what this bound gives us, making it an upper bound. However, this does not mean that due to a small phase register, or coincidental phase interference, or because of finite sampling, the optimal phase bitstring can never be measured more frequently than $p(x)$, therefore, this is not a hard upper bound but an \emph{effective upper bound}. Importantly this ``bound'' in Eq.~\eqref{equation:px_upper_bound} is additionally not a standard bound in the sense that the value cannot be computed for the optimal phase (e.g., before the QPE computation is run) because that requires knowing the minimum eigenvalue -- we only compute it here because we can easily compute the minimum eigenvalue because the Hamiltonians are small. The minimum value of the optimal phase sampling $p(x)$ in Eq.~\eqref{equation:px_upper_bound} is $\frac{4}{\pi^2}$, which occurs if the eigenphase is exactly in between two phase bitstrings.

\begin{figure}[ht!]
    \centering
    \includegraphics[width=1.0\linewidth]{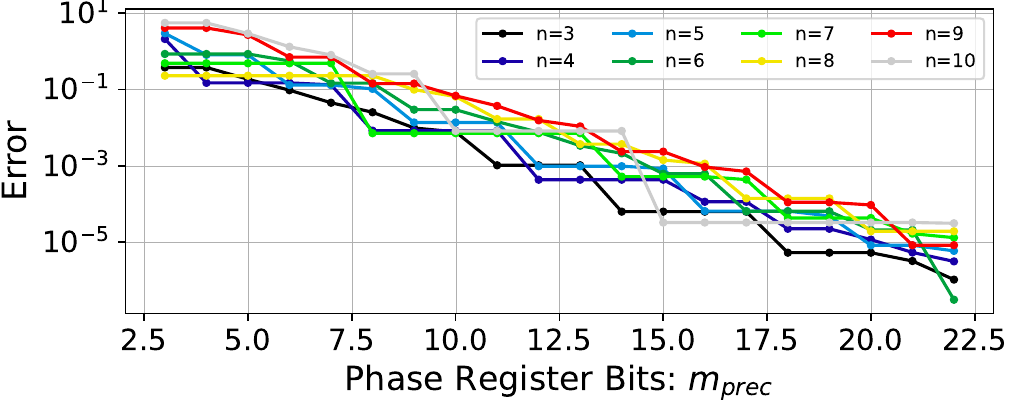}
    \caption{ \textbf{Digitization phase readout error.} Error (log-scale) with respect to the optimal quantum Hamiltonian ground-state energy as a function of increased bits provided for the digitization approximation of the ground-state energy, using a total time evolution given by Eq.~\eqref{equation:heuristic_evolution_time} and the minimum eigenvalue found by exact diagonalization for Heisenberg spin glass models with up to $10$ qubits. Error is defined as the absolute value of the difference between the ground state energy ($E_0$) and the closest bitstring digitization of that energy. Note that this data is not from a QPE circuit simulation, rather, this is showing the lowest possible error for each number of bits of precision up to $22$.  }
    \label{fig:phase_bit_error}
\end{figure}

\begin{figure}[ht!]
    \centering
    \includegraphics[width=1.0\linewidth]{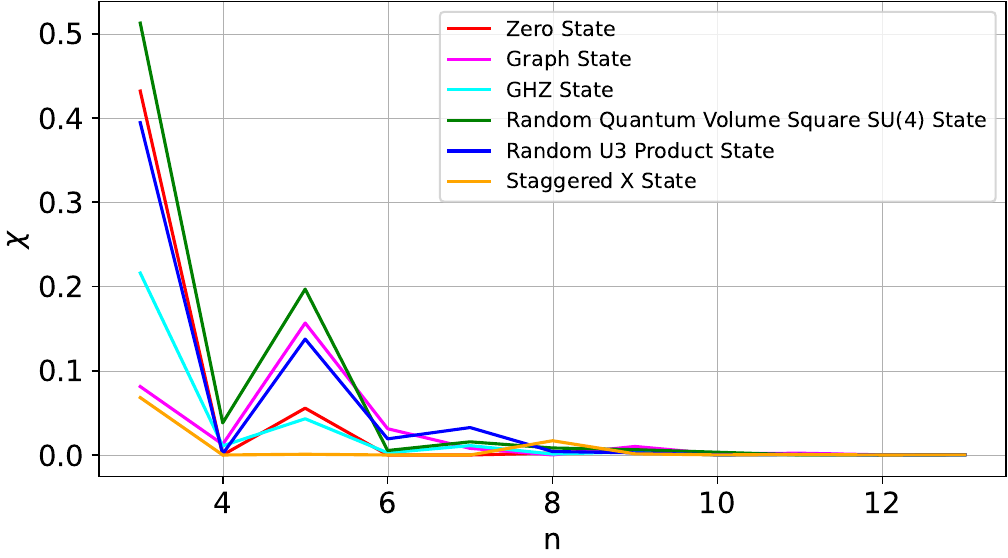}
    \caption{ Overlap (y-axis) between various easy-to-prepare initial states and the ground-states(s) eigenvector(s) of the quantum Hamiltonian (x-axis). The overlap quantity $\chi$ is defined by Eq.~\eqref{equation:degenerate_overlap}. }
    \label{fig:overlap}
\end{figure}

\paragraph*{Steady-state optimal digitized phase sampling proportion.}Taking into account both initial state overlap and the time-periodicity of the optimal phase, the actual optimal phase sampling ``steady-state'' probability is 

\begin{equation}
    \label{equation:QPE_optimal_phase_sampling_rate}
    \zeta = \chi \cdot p(x). 
\end{equation}

\begin{figure*}[ht!]
    \centering
    \includegraphics[width=0.495\linewidth]{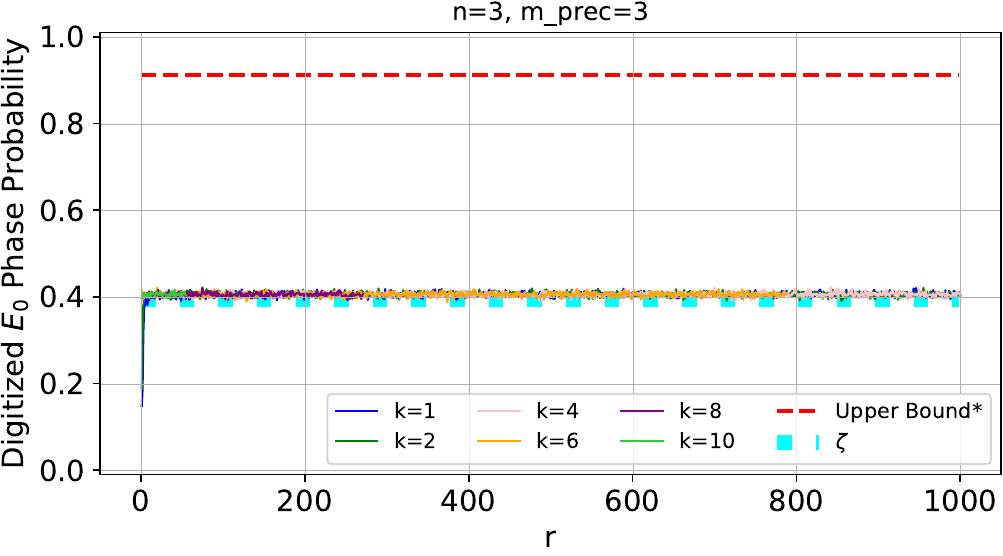}
    \includegraphics[width=0.495\linewidth]{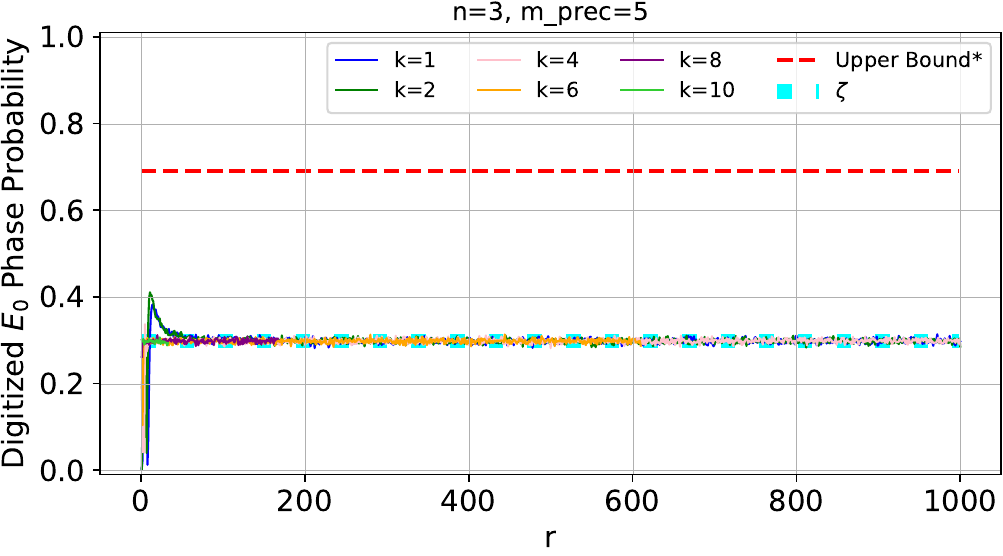}
    \includegraphics[width=0.495\linewidth]{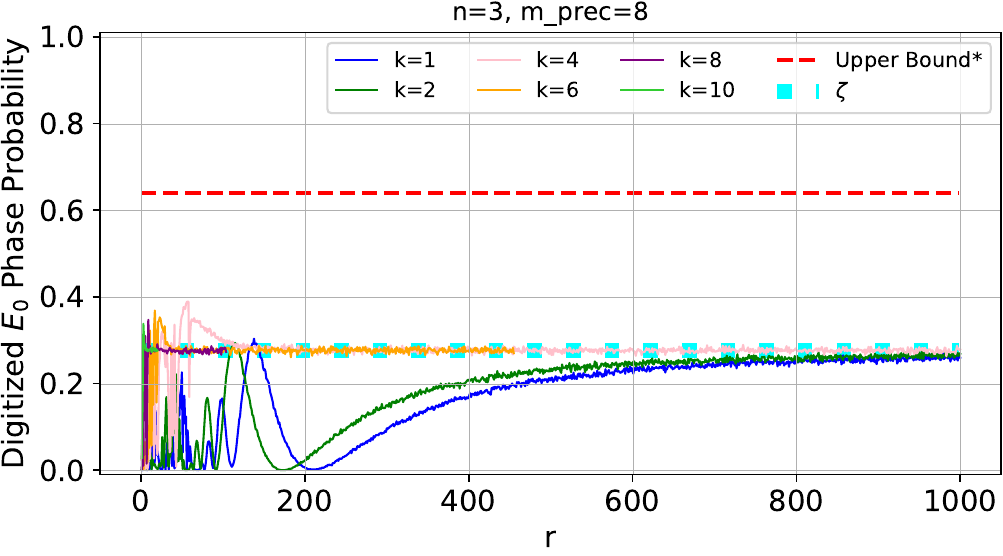}
    \includegraphics[width=0.495\linewidth]{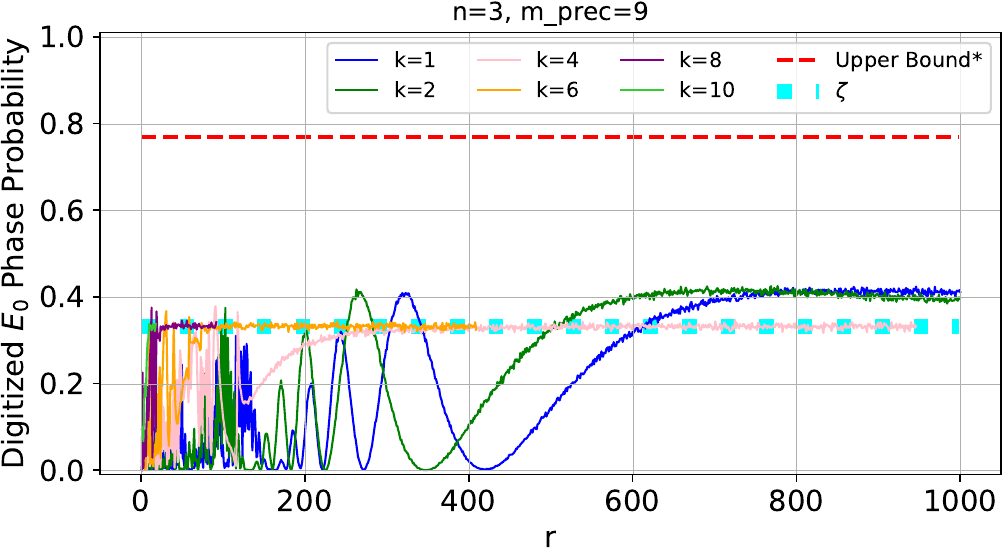}
    \caption{ \textbf{ QPE optimal phase sampling (y-axis) as a function of Trotter steps $r$ (x-axis)}. The time evolution used is the $t_0$ from Eq.~\eqref{equation:heuristic_evolution_time}. At small $r$, we observe clear transitory effects due to the time evolution being approximated poorly. Higher order $k$ at larger $r$ is not reported because of the substantial gate counts required to represent the full QPE circuit, and more specifically the classical circuit simulation of those large circuits. The dashed red line labeled \emph{Upper Bound*} plots Eq.~\ref{equation:px_upper_bound}, and the dashed cyan line plots Eq.~\ref{equation:QPE_optimal_phase_sampling_rate}.  }
    \label{figure:function_of_r}
\end{figure*}

Recall that $\chi$ is the initial state overlap. The quantity $\zeta$ is defined for a specific Hamiltonian instance (which we must exactly diagonalize to obtain all degenerate eigenvalues and the corresponding eigenvectors), a total evolution time $t$, and an initial state $\mathcal{A}$. $\zeta$ is the steady-state optimal phase sampling of QPE, given that the Hamiltonian evolution is approximated sufficiently well. Note that $\zeta$ does account for phase leakage from other eigenvalues; a full statespace probability distribution could be obtained, including phase leakage, by enumerating over all $2^n$ possible phases using Eq.~\eqref{equation:px_upper_bound} however Eq.~\ref{equation:QPE_optimal_phase_sampling_rate} is easier to compute particularly when the number of phase bits is large. Such leakage typically becomes apparent at small phase register sizes. $\zeta$ is a very useful diagnostic for QPE implementation, at least for scales at which $\zeta$ can be verified using exact diagonalization.

\paragraph*{Trotter error} To measure Trotter error, we compute the Frobenius norm between the ideal unitary evolution, at the specified evolution time, and the unitary created by the Trotterized circuit 

\begin{equation}
\label{equation:trotter_error_frobenius_norm}
|| U_{\text{ideal}} - U_{\text{Trotterized}} ||_F. 
\end{equation}

Note that this is not Trotter error within QPE, necessarily, because QPE includes multiple controlled unitaries.

\begin{figure*}[ht!]
    \centering
    \includegraphics[width=0.495\linewidth]{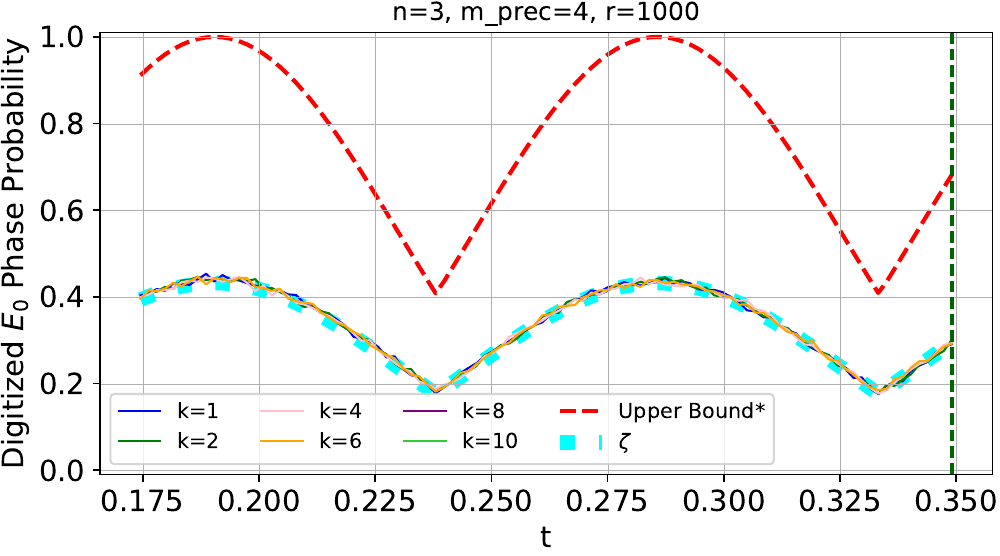}
    \includegraphics[width=0.495\linewidth]{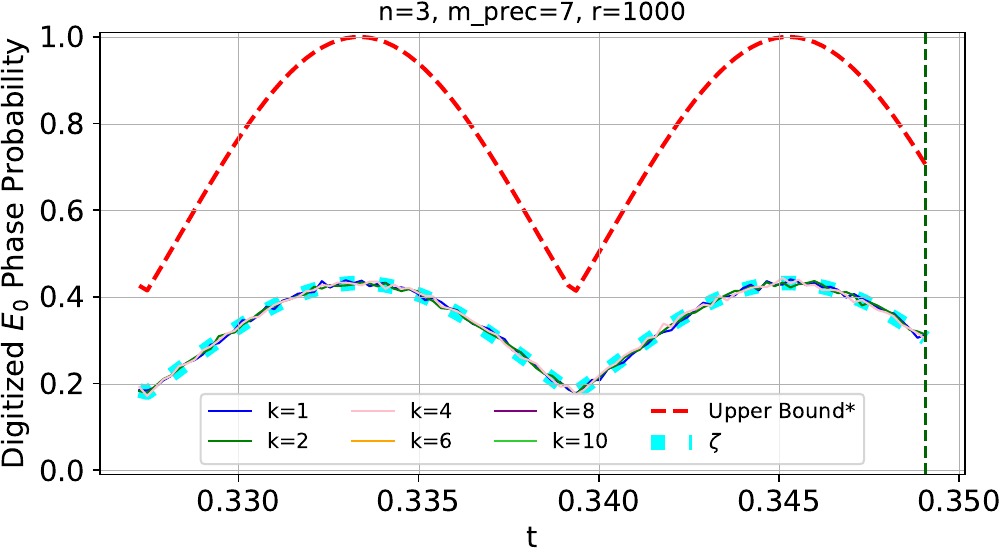}
    \includegraphics[width=0.495\linewidth]{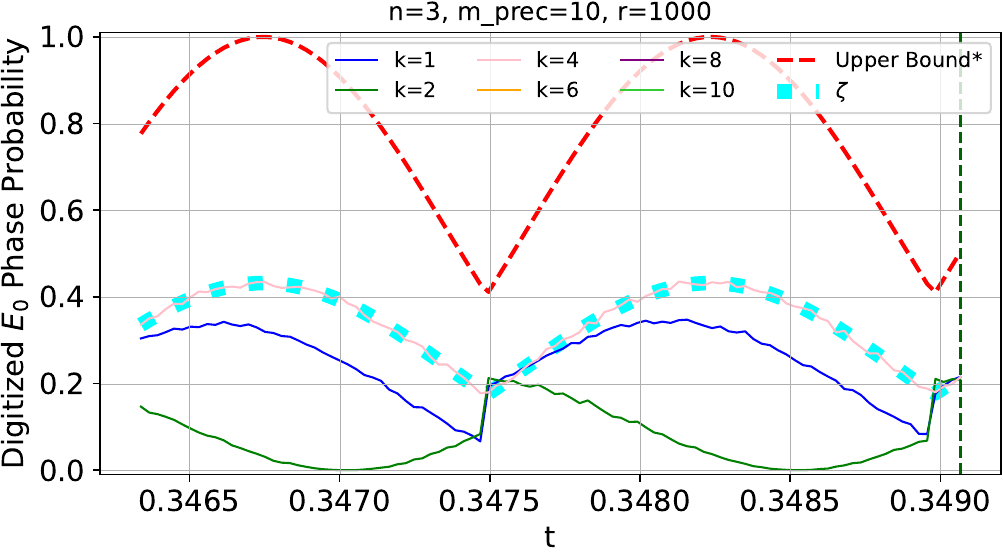}
    \includegraphics[width=0.495\linewidth]{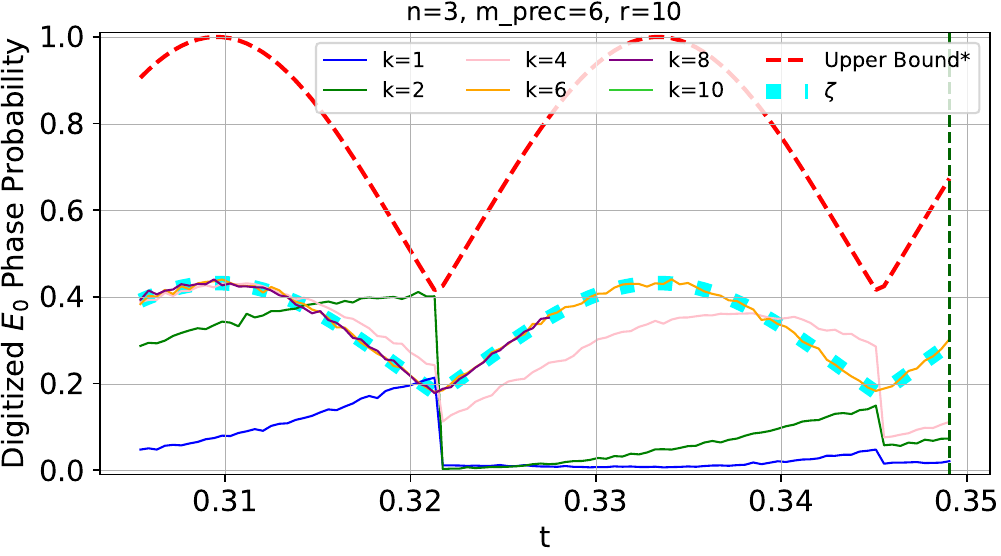}
    \caption{ \textbf{High resolution search over evolution times.} Measuring the optimal phase sampling rate over a gridsearch of evolution times specified by a range starting from the time given by Eq.~\eqref{equation:heuristic_evolution_time} (vertical dashed green line), and then down to $t_0 - 8\cdot \frac{t_0}{2^{m_{prec}}}$. This scaling of the time resolution is because the more phase qubits are used, the shorter the periodicity of the time evolution becomes. The dashed red line labeled \emph{Upper Bound*} plots Eq.~\ref{equation:px_upper_bound}, and the dashed cyan line plots Eq.~\ref{equation:QPE_optimal_phase_sampling_rate} as a function of time. In all four sub-plots the time evolution begins at precisely the same $t_0$, however, \emph{where} in this periodic time landscape $t_0$ is depends on $m_{prec}$. Note that when $t$ changes, the optimal phase can also change, which means that the y-axis is not necessarily plotting the sampling rate of a single bitstring. The bottom row shows two examples of the transitory effects that occur when the Trotter error is too high, and the therefore the sampling rate has not converged to Eq.~\eqref{equation:QPE_optimal_phase_sampling_rate}. 
    }
    \label{fig:function_of_time_high_resolution}
\end{figure*}

\begin{figure*}[ht!]
    \centering
    \includegraphics[width=0.495\linewidth]{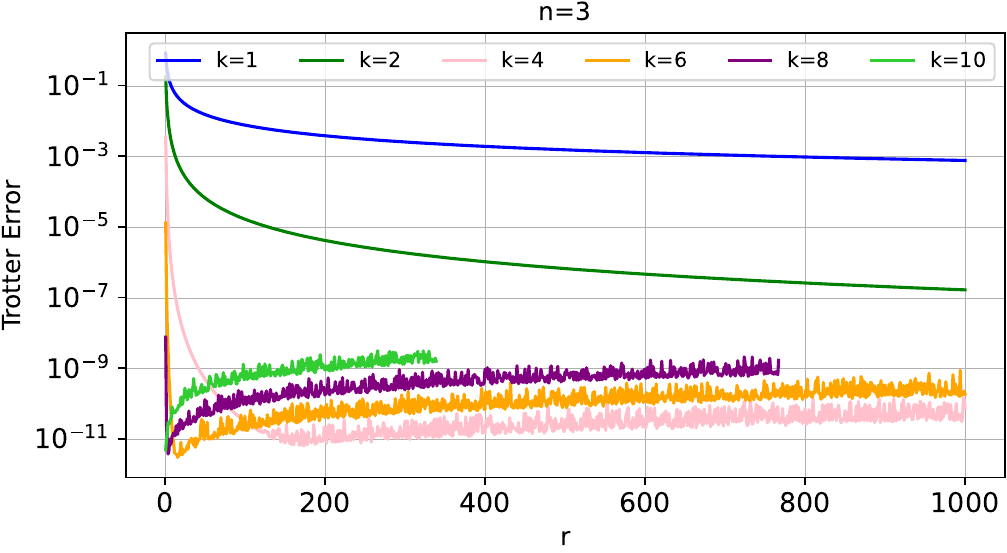}
    \includegraphics[width=0.495\linewidth]{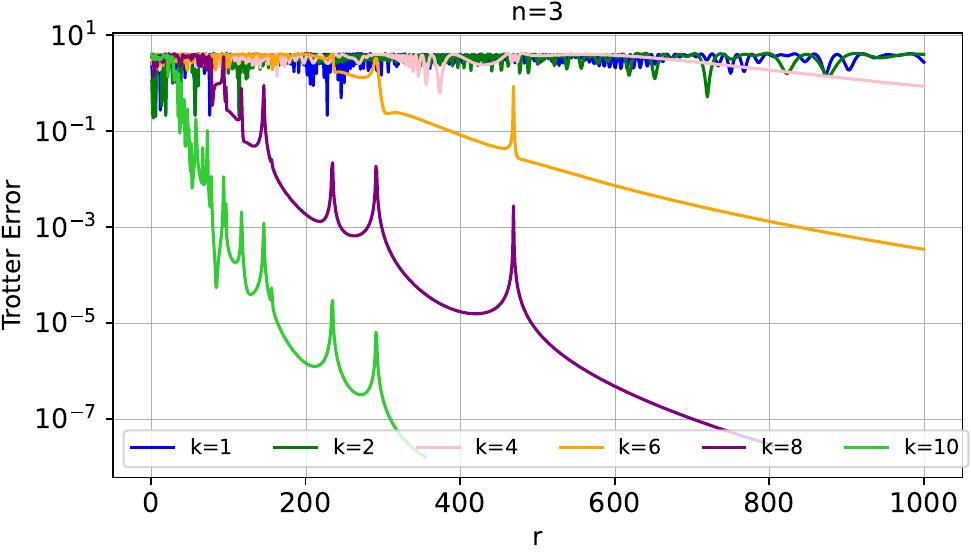}
    \caption{  \textbf{Trotter error, in terms of the Frobenius norm, as a function of Trotter steps ($r$) and Trotter order ($k$).} Trotter error (y-axis in Eq.~\eqref{equation:trotter_error_frobenius_norm} as a function of $r$ (x-axis) for a $n=3$ Hamiltonian. The left sub-plot shows the error from approximating $e^{i H t_0 2^0}$, which corresponds to the time evolution required for the first phase qubit in the QPE circuit. The right sub-plot shows the error when approximating $e^{i \mathcal{H} t_0 2^{10}}$, which corresponds to the time evolution required for the $11$th phase qubit in a (up to) $11$ phase qubit register QPE circuit. $t_0$ is given by Eq.~\eqref{equation:heuristic_evolution_time}.  }
    \label{fig:Trotter_error}
\end{figure*}

\section{Results}
\label{section:results}

The first most fundamental source of error from QPE, as with all digital algorithms, is the digitization of the true eigenvalue number. Fig.~\ref{fig:phase_bit_error} shows the relative error of the true ground-state energy as a function of the number of bits used in the phase register, for quantum Hamiltonians of various sizes, using the total time evolution given by the generic Hamiltonian coefficient summation bound of Eq.~\eqref{equation:heuristic_evolution_time}. For a given phase register length, we will only ever be able to get as good of an error as that digitization allows, and therefore for the remainder of the plots and discussion we will address the question of sampling rate of the \emph{optimal} phase register (which corresponds to the minimum eigenvalue), given a fixed $m_{prec}$. 

The next most important input to QPE is the initial state $\mathcal{A}$, specifically the overlap quantity defined in Eq.~\eqref{equation:degenerate_overlap}. For the purposes of numerically running QPE circuits, we must choose some reasonable initial states that have good overlap. To this end, Fig.~\ref{fig:overlap} plots the overlap quantity as a function of quantum Hamiltonian size $n$ (with one Hamiltonian instance per $n$), for several different easy-to-prepare initial states. In the literature, as far as we are aware there is very limited study on high-overlap initial states for Heisenberg spin glass models. For the initial states we considered, the overlap was not very large, and for larger $n$ the overlap decreases significantly. Entangled states like the GHZ state, random SU(4) square ``quantum volume'' states, and graph states, all did not have very good overlap. The all zero state for $n=3$ and $n=4$ had reasonably good overlap, and therefore for the remainder of the text we will focus $n=3$ and the all zero initial state. The all zero state is not very representative, and is not expected to be a very good initial state for very large system sizes, however, it serves the purposes of numerical parameter experimentation because it happens to have high overlap for very small cases, and is extremely easy to prepare. Appendix~\ref{section:appendix_averaged_overlap} reports initial state overlap measures averaged over many realizations of the Heisenberg spin glass model. 
The lack of high-overlap initial states for these types of models points to the relatively well understood viewpoint that much of the algorithmic engineering work of QPE will be initial state preparation.

The next source of error to consider is how well Trotterization approximates the time evolution. Fig.~\ref{figure:function_of_r} plots the optimal phase sampling rate from a full QPE circuit, using $t_0$ from Eq.~\ref{equation:heuristic_evolution_time}, as a function of both number of Trotter steps $r$ and Trotter order $k$, which show that QPE quickly converges to the optimal phase sampling rate defined by Eq.~\ref{equation:QPE_optimal_phase_sampling_rate}. Next, we show how evolution time is a crucial parameter as well; Fig.~\ref{fig:function_of_time_high_resolution} shows how the optimal phase sampling rate changes as the evolution time is changed. For all of these elementary quantum gate QPE circuit numerical simulations, we focus on the smallest Heisenberg model with $3$ qubits because this reduces total gate count overhead for simulating the circuits, and the all zero initial state overlap (Fig.~\ref{fig:overlap}) is large enough that the optimal phase sampling rate is visually apparent. 
This shows that the optimal phase sampling rate changes significantly as the evolution time is varied -- for a perfect initial state (shown by the dashed red line) this sampling rate changes from $\frac{4}{pi^2} \approx 0.4$ to $1.0$. The periodicity of the time evolution increases exponentially as $m_{prec}$ increases. Crucially, both Fig.~\ref{figure:function_of_r} and Fig.~\ref{fig:function_of_time_high_resolution} show that the QPE performance, specifically in terms of optimal sampling rate, adheres to the steady-state sampling rate given by Eq.~\ref{equation:QPE_optimal_phase_sampling_rate}. Fig.~\ref{fig:Trotter_error} plots Trotter error (without accounting for the controls that would be used in the QPE circuit) as a function of both $r$ and $k$ for two different total evolution times. The Trotter error context is notable because both Fig.~\ref{figure:function_of_r} and Fig.~\ref{fig:function_of_time_high_resolution} show that there is a clear diminishing return in QPE performance (in terms of optimal phase sampling rate) as Trotter error is decreased. In other words, practically it is not always necessary to push the Trotter error to be as low as possible. This shows that a fairly high Trotter error, especially when $m_{\text{prec}}$ is small, can still result in sampling at the optimal phase steady-state. The noise floor in Fig.~\ref{fig:Trotter_error} is determined mainly by the floating point precision limits of the classical numerical computations. The transitory Trotter error in Fig.~\ref{fig:Trotter_error} is notable because the error is not necessarily monotonic with respect to $r$ for a given $k$, or with respect to $k$ for a given $r$.

\begin{figure*}[ht!]
    \centering
    \includegraphics[width=0.495\linewidth]{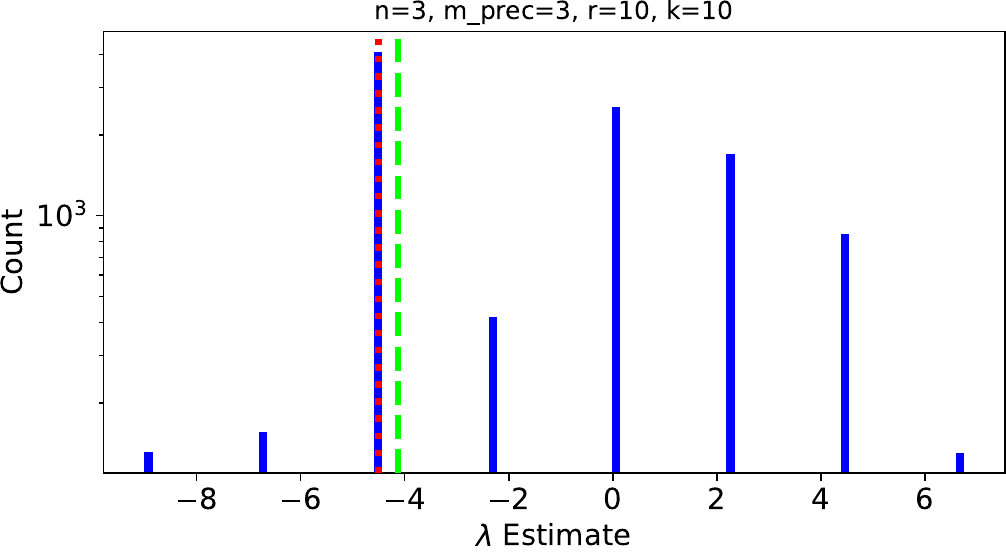}
    \includegraphics[width=0.495\linewidth]{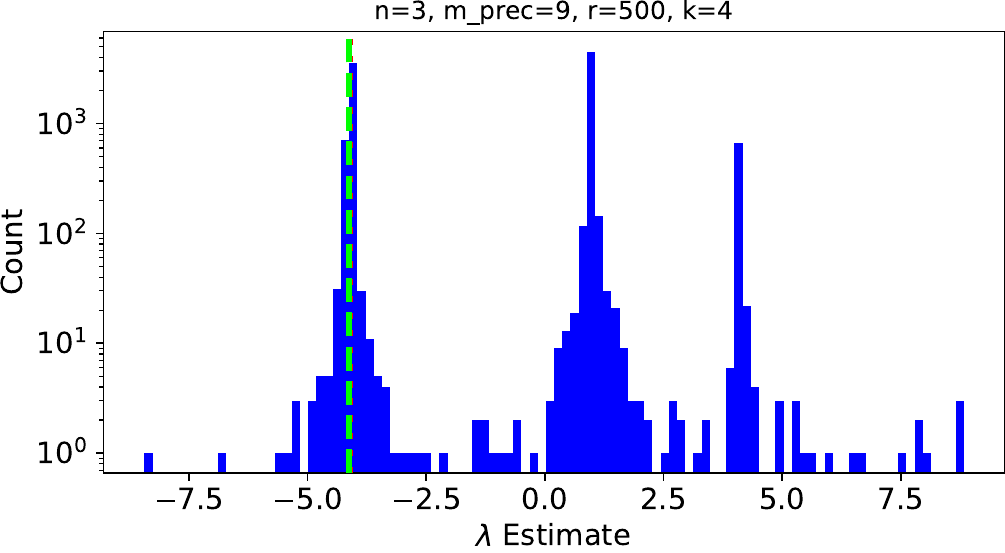}
    \caption{ \textbf{Comparing low bit precision ($3$) and high bit precision ($9$) decoded QPE eigenvalue estimates. } Vertical dashed red line is the optimal eigenvalue digitized energy (at this fixed time evolution, for each Hamiltonian instance). Vertical lime green dashed line is the true eigenvalue; the distance between the red line and the lime green line is the digitization error. In both cases, the Trotter parameters $r, k$ here mean that the QPE sampling converges to the optimal phase sampling rate of Eq.~\eqref{equation:QPE_optimal_phase_sampling_rate}, as shown in Fig.~\ref{figure:function_of_r}. Note that all eigenvalue estimates to the left of the green/red vertical lines are \emph{non-physical} eigenvalue estimates due to the tail distributions inherent to this QPE distribution, regardless of how small the Trotter error is. The y-axis count is log scale.   }
    \label{fig:energy_distributions_low_high_bit_precision}
\end{figure*}

\begin{figure*}[ht!]
    \centering
    \includegraphics[width=0.495\linewidth]{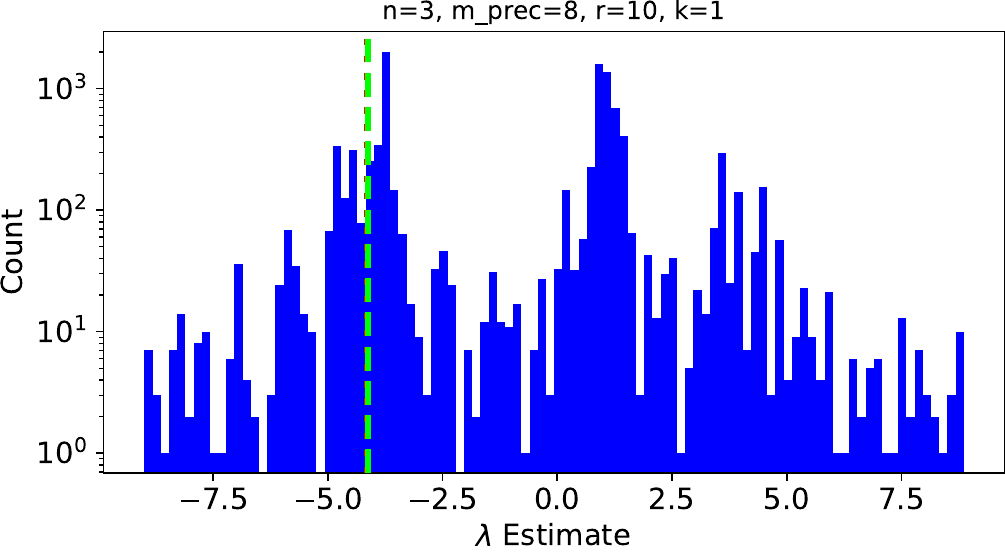}
    \includegraphics[width=0.495\linewidth]{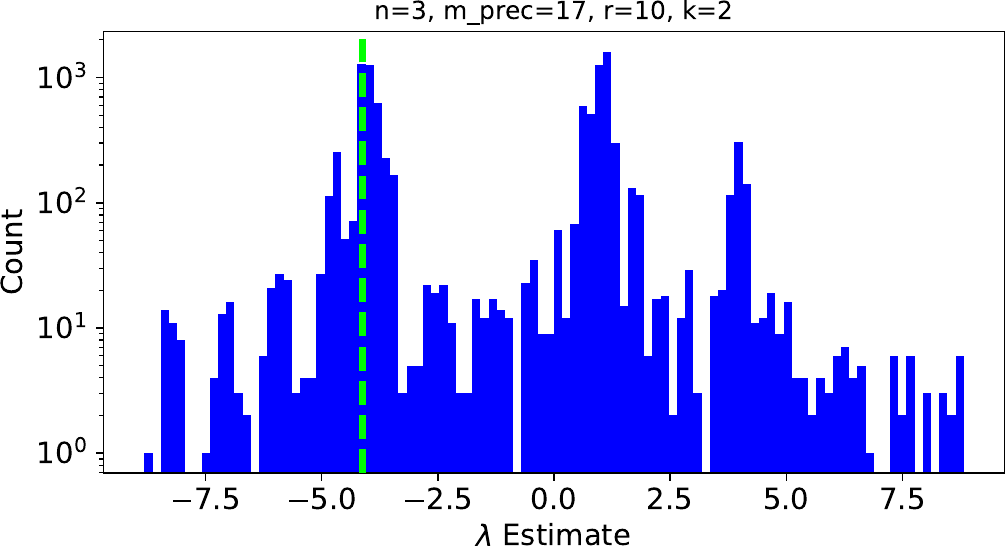}
    \caption{ \textbf{QPE decoded eigenvalue estimates where the Trotter error is substantially high, resulting in QPE sampling that has not converged to the optimal phase sampling rate.} This shows that the eigenvalue estimate distributions that we get when the approximated unitary $\tilde{U}$ is not a good approximation of $U$ are not very concentrated and moreover there is an even higher probability of non-physical eigenvalue estimates compared to Fig.~\ref{fig:energy_distributions_low_high_bit_precision}.  }
    \label{fig:energy_distributions_bad_unitary_approximation}
\end{figure*}

Next we examine the full estimated eigenvalue distributions; Fig.~\ref{fig:energy_distributions_low_high_bit_precision} shows distributions from the highest ($9$) and lowest ($3$) bit phase precision we have tested, where the QPE sampling has converged due to sufficiently low Trotter error. The full eigenvalue distributions show a crucial aspect of standard QPE, which is that some of the estimated eigenvalues can be smaller than the actual minimum eigenvalue $E_0$. Fig.~\ref{fig:energy_distributions_bad_unitary_approximation} shows eigenvalue estimates when the Trotter error is quite high, resulting in the digitized $E_0$ phase not having converged to Eq.~\ref{equation:QPE_optimal_phase_sampling_rate}. 

\begin{figure*}[ht!]
    \centering
    \includegraphics[width=0.495\linewidth]{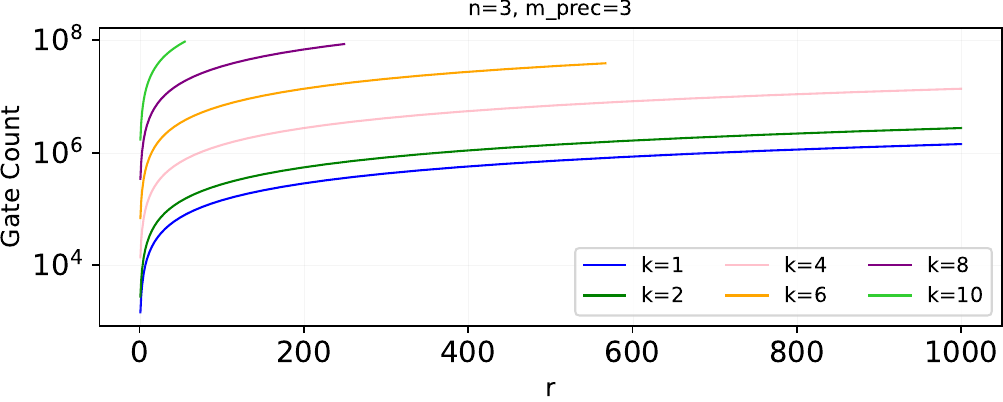}
    \includegraphics[width=0.495\linewidth]{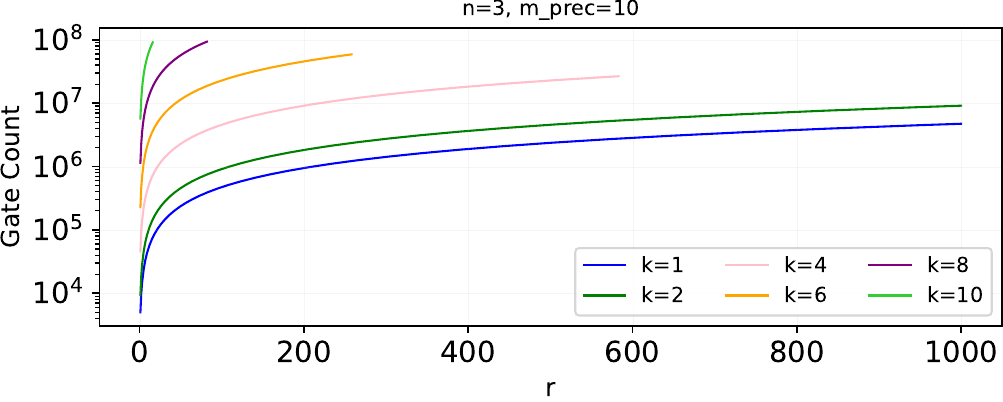}
    \caption{ \textbf{ \textbf{Total QPE circuit instruction count (log-scale y-axis is the sum of 1Q and 2Q gates) as Trotter steps and Trotter order are increased. } }  }
    \label{fig:QPE_circuit_count}
\end{figure*}

Finally, Fig.~\ref{fig:QPE_circuit_count} reports the total single- and two-qubit gate count of the full QPE circuits as Trotter step and Trotter order increase. This shows the exponential increase in total gate count that becomes apparent for higher Trotter orders. Note that these are entirely unoptimized QPE circuits.

\section{Discussion and Conclusion}
\label{section:conclusion}

The numerical experiments that we have reported are some of the only examples that have been presented in the literature of actual QPE circuit execution (or numerical simulations) aimed at ground-state computation of quantum Hamiltonians, including tuning of input parameters (namely, time, Trotter order, and Trotter steps). There are two clear well-known drivers of continued QPE algorithmic improvement; better initial state preparation, and more efficient time dynamics simulation. The primary ways in which the time dynamics simulation can be practically improved is by i) optimizing existing product formulas such as permuting term reordering in the Trotter product formulas~\cite{Childs_2021, Childs_2019, Tranter_2019, hastings2014improvingquantumalgorithmsquantum}, ii) using new optimized higher order Trotter-Suzuki schemes~\cite{maležič2026efficienttrottersuzukischemeslongtime, maležič2026reducinggatecountefficient, Ostmeyer_2023}, and iii) using time dynamics compression techniques based on tensor network methods~\cite{Gibbs_2025, gibbs2025learningcircuitsinfinitetensor}.

Next we outline the key findings that these small scale numerical experiments have elucidated for standard QPE. Importantly, these properties are not novel algorithmic properties -- they are known from the theoretical properties of QPE -- however these properties are, to the best of our knowledge, not well-known to the quantum algorithms community, and are consequential for the actual operation of QPE. Many of these characteristics we describe only become apparent when QPE is actually run, either on a quantum computer or with brute-force classical statevector numerical computations. Some of these effects are transitory effects due to poor approximation of the unitary evolution and finite system sizes.

\emph{1. Using the smallest energy found from the measured QPE circuit executions can result in an energy that is not physical, i.e., the energy is lower than the actual ground-state of the quantum Hamiltonian (see Figs.~\ref{fig:energy_distributions_low_high_bit_precision},~\ref{fig:energy_distributions_bad_unitary_approximation}).} This is a counter-intuitive property -- especially from the context of standard classical linear algebra computation. This known property is due to the probability distribution generated by QPE~\cite{PRXQuantum.5.040339}, in particular when the digitized phase is not a good approximation of the true eigenvalue. Informally, this can be viewed as a quantum equivalent of the decay tails in the classical Fourier transform. This feature has spurred subsequent algorithmic improvements to the standard QPE, namely a coherent median computation technique~\cite{nagaj2009fastamplificationqma}, and tapered QPE~\cite{patel2024optimalcoherentquantumphase}. In addition to algorithmic improvements, there are computationally cheap methods that can be employed in post-processing of standard QPE outputs, such as efficient eigenvalue bounds to reject some output energies as non-physical. These non-physical eigenvalue estimates do not occur in perfectly idealized QPE if 1) the initial state has perfect overlap, 2) the time evolution is tuned to be at a peak of Eq.~\eqref{equation:px_upper_bound}, and 3) the Trotterization is a good approximation of the true time evolution.

\emph{2. The total evolution time is an important parameter that can be tuned in order for QPE to perform especially well.} Eigenvalue bounds can be used in order to obtain reasonably good total evolution time, however, the total evolution time can be tuned to generate higher optimal phase sampling rates compared to what an initial time evolution bound can produce. Moreover, the time evolution changes the distribution of eigenvalues that are sampled. For a given fixed initial state with some maximum overlap $x$, the difference between no tuning of the time evolution and finding an optimal evolution time is a factor of $4/\pi^2\approx0.4x$ compared to $1x$, as shown in Fig.~\ref{fig:function_of_time_high_resolution}.

\emph{3. There is a saturation, or a steady-state, of the optimal phase sampling that occurs at sufficiently low error rate.} This property of QPE is good; it means that there does exist some sufficient error rate at which the QPE sampling converges, which in particular means that Trotter error does not need to be decreased \emph{ad infinitum}. This is illustrated by Fig.~\ref{figure:function_of_r} and Fig.~\ref{fig:Trotter_error}. This property is not necessarily apparent given the scaling, error, or asymptotic properties of QPE. The existence of the steady state at still relatively high Trotter error (compared to for example, standard machine precision) is useful to know of for diagnosing QPE implementations and bounding expected large-scale QPE performance. The sampling rate is explicitly given by Eq.~\eqref{equation:QPE_optimal_phase_sampling_rate}. Unfortunately, computing both of the quantities in Eq.~\eqref{equation:QPE_optimal_phase_sampling_rate}, and therefore the overall ground-truth optimal phase QPE sampling rate cannot be computed easily, in fact in both cases this requires knowing the minimum eigenvalue of the Hamiltonian. So, this quantity is a useful diagnostic for small-scale QPE circuit implementation and testing, as well as theoretical analysis of QPE, but its direct computation will likely not be useful for large scale QPE implementation on fault tolerant quantum computers.

\emph{4. Perfect QPE, run on a noiseless quantum computer,  with a good initial state would not always output the optimal phase, corresponding to ground-state energy.} This is in part due to the probabilistic nature of the computation, but moreover the probability distributions are not guaranteed to concentrate on the optimal phase. The optimal phase sampling rate is determined solely by the initial state overlap and the evolution time. If the initial state is not perfect, which it will never be in practice, then eigenvalue estimates can be produced within the eigenvalue spectrum of the Hamiltonian (e.g., see Figs.~\ref{fig:energy_distributions_low_high_bit_precision} and \ref{fig:energy_distributions_bad_unitary_approximation}). And if the time evolution is not a peak of Eq.~\eqref{equation:px_upper_bound}, then the optimal phase sampling rate will not always be sampled even if the initial state is perfect, see Figs.~\ref{figure:function_of_r} and~\ref{fig:function_of_time_high_resolution}. This property is inherent to the algorithm, so it is not an operational issue with the algorithm, but the overall success probability could be boosted by amplitude amplification~\cite{rajchelmieldzioć2026quantumalgorithmsolvinggeneralized, Brassard_2002, Daskin_2018, grover1996fastquantummechanicalalgorithm}.

\emph{5. Transitory effects due to poor approximation of the time evolution can cause outlier behavior. For instance, higher success probability than steady-state converged QPE value can occur (see Fig.~\ref{figure:function_of_r}).} This is not necessarily a problem. In fact, given that in some instances it boosts success probability to be greater than Eq.~\ref{equation:QPE_optimal_phase_sampling_rate} it may be able to be exploited to improve QPE in some way. Fig.~\ref{fig:function_of_time_high_resolution} illustrates transitory effects that occur as the evolution time causes the optimal phase bitstring to switch over, where the optimal phase sampling reverses from high to low, or vice versa.

\emph{6. Phase leakage can cause slightly higher digitized $E_0$ eigenvalue steady-state sampling probability of Eq.~\eqref{equation:QPE_optimal_phase_sampling_rate}.} This occurs because of leakage from nearby eigenvalues to $E_0$, and becomes more apparent when $m_{prec}$ is small. This is illustrated in Fig.~\ref{figure:function_of_r}-(top left) with $3$ phase qubits, although the effect in this plot is relatively small.

\section*{Acknowledgments}
\label{sec:acknowledgments}
The authors thank Yigit Subasi, Andreas Bärtschi, John Golden, Sumner Hearth, and Abhijith Jayakumar for many productive discussions. 
This work was supported by the U.S. Department of Energy through the Los Alamos National Laboratory. Los Alamos National Laboratory is operated by Triad National Security, LLC, for the National Nuclear Security Administration of U.S. Department of Energy (Contract No. 89233218CNA000001). This research used resources provided by the Los Alamos National Laboratory Institutional Computing Program. This research used resources provided by the Darwin testbed at Los Alamos National Laboratory (LANL) which is funded by the Computational Systems and Software Environments subprogram of LANL's Advanced Simulation and Computing program (NNSA/DOE). This research used resources provided by the Los Alamos National Laboratory Institutional Computing Program, which is supported by the U.S. Department of Energy National Nuclear Security Administration under Contract No.~89233218CNA000001. Research presented in this article was supported by the NNSA's Advanced Simulation and Computing Beyond Moore's Law Program at Los Alamos National Laboratory. LANL report LA-UR-26-20498.

\appendix

\section{QPE Qiskit Circuit Implementation and Design Considerations}
\label{section:appendix_QPE_circuit_implementation}

There are at least three different ways to implement the Trotterized controlled unitary evolution within QPE, all three of which we briefly describe here using the notation of Fig.~\ref{figure:QPE_circuit}. 

\begin{figure}[ht!]
    \centering
    \includegraphics[width=1.0\linewidth]{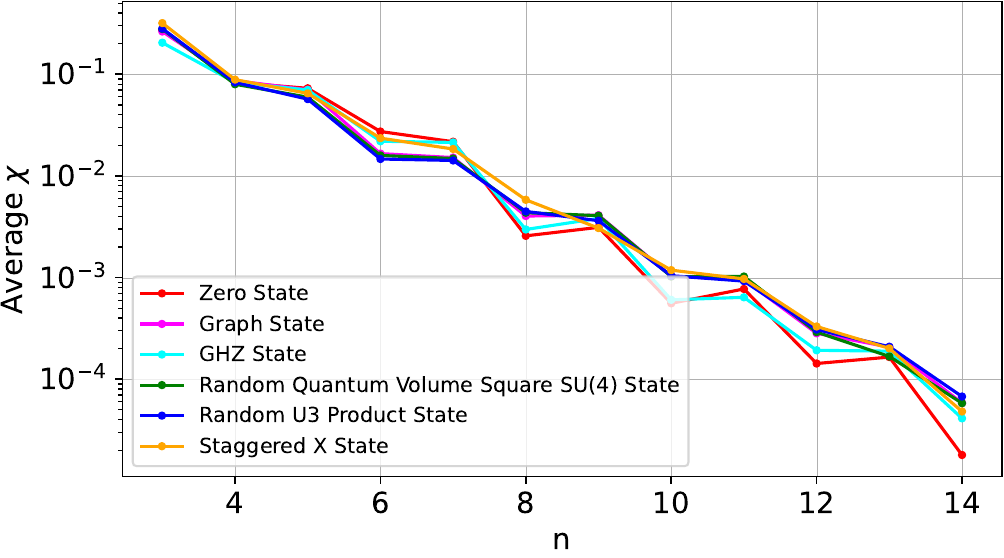}
    \caption{ Averaged initial state overlap over $100$ random instances per $n$, with several different easy to prepare states. Log-scale y-axis. None of the evaluated states provide a substantially higher overlap than any of the others, and as a function of $n$ the overlap decreases exponentially.  }
    \label{fig:averaged_overlap}
\end{figure}

The first is the most direct and naive; for each controlled unitary evolution, we repeat the same (Trotterized) unitary evolution for time $t$, but repeated $2^j$ times where $j$ indexes the phase qubit register from $0$ up to $j=m_{prec}$. This version has an exponential controlled unitary evolution count with the number of bits of precision, which results in very high depth circuits when the number of bits of precision is high. This version should never be used in practice, although it is technically correct.

The second way is by casting the Trotterization approximation to a full unitary, and then carrying out full matrix exponentiation of the unitary, by $2^j$ where $j$ indexes the phase qubit register. Each controlled unitary is applied only once per phase qubit. This method works well, but is not scalable and requires matrix exponentiation by the total evolution time. This requires then enforcing strict unitarity conditions which primarily requires handling floating point error accumulation in the unitary matrix exponentiation stage causing the operator to become non-unitary. This can be fixed by casting all integer-exponentiated unitary matrices to the closest true unitary using Singular Value Decomposition (SVD) and high-precision floating point operations. Different SVD approaches may be required, namely \texttt{gesdd} and \texttt{gesvd} from LAPACK~\cite{129995}, as well as a recursive SVD calls on each subsequent component matrix. If SVD attempts failed, polar decomposition~\cite{horn2012matrix} from scipy~\cite{2020SciPy-NMeth} can be utilized. Note that this particular numerical precision issue only becomes apparent at high Trotter step sizes and high Trotter orders ($4$ or greater).

The third way is the correct and scalable approach, which is that each controlled time evolution is run for a time of $t \cdot 2^j$ (each Trotter decomposition uses this different time evolution for each controlled unitary operation), again where $j$ indexes the phase qubit register. Once again, each controlled unitary is applied only once per phase qubit.

These three implementations are equivalent unitarily, but can have different circuit level decompositions. For the Qiskit implementation we used the third option, where time is exponentiated separately for each phase qubit. Quantum circuit compilation is another aspect of QPE circuits which we do not consider; all circuits are adapted into a general single- and two-qubit gateset using Qiskit~\cite{javadiabhari2024quantumcomputingqiskit}. Crucially, the decomposed QPE circuits, even for relatively small $r$ and $k$ are so large that compiler optimization, such as in Qiskit, requires such a significant amount of compute time that we considered this intractable using current quantum circuit compiler software with any reasonable level of classical compute.

\section{Averaged Initial State Overlap}
\label{section:appendix_averaged_overlap}

Fig.~\ref{fig:averaged_overlap} shows initial state overlap, averaged over $100$ random coefficient instantiations of Eq.~\eqref{eq:heisenberg_glass}, as a function of $n$. This shows that none of the initial states have very high overlap, and the overlap is decreasing exponentially as a function of $n$. Significantly improved initial state methods must be developed in order for QPE to be applied to Heisenberg quantum Hamiltonian ground-state computation, beyond these simple test cases.

\clearpage

\bibliographystyle{apsrev4-2-titles}
\bibliography{references}
\end{document}